\documentclass[%
superscriptaddress,
reprint,
 amsmath,amssymb,
 aps,
prb,
]{revtex4-2}

\usepackage{graphicx}
\usepackage{dcolumn}
\usepackage{bm}
\usepackage{natbib} 
\usepackage{color}
\usepackage{subfigure}
\usepackage{siunitx} 

\begin{document}

\newcommand{\comment}[1]{{\color{red}#1}}

\newcommand{\ueff}[0]{$U_{\mathrm{eff}}$}

\preprint{APS/123-QED}

\title{Magnetic, Structural, and Electronic Properties of CrOCl: Insights from the PBE Functional}  

\title{Magnetic, Structural, and Electronic Properties of CrOCl with the PBE Functional}

\newcommand{\ifm}{Theoretical Physics Division, Department of Physics, Chemistry and Biology (IFM), Link\"oping University, SE-581 83 Link\"oping, Sweden}

\author{Brahim Marfoua}\email{brahim.marfoua@liu.se} 
\affiliation{\ifm}

\author{Mohammad Amirabbasi} 
\affiliation{Technische Universit\"at Darmstadt, Fachbereich Material und Geowissenschaften, Fachgebiet Materialmodellierung, Otto‑Berndt‑Straße 3, 64283 Darmstadt, Germany}

\author{Marcus Ekholm}\email{marcus.ekholm@liu.se}
\affiliation{\ifm}

\begin{abstract}
CrOCl is a van der Waals–layered insulator with an antiferromagnetic ground state, making it a promising platform for exfoliation and the exploration of low-dimensional magnetism. An accurate ab initio description is therefore essential. Previous density-functional studies have shown that DFT+$U$ calculations may erroneously favor ferromagnetic order depending on the choice of parametrization, an issue that cannot be remedied by simply adjusting the value of $U$. Here, we demonstrate that an explicit Hubbard correction is unnecessary: the PBE functional correctly reproduces the AFM ground state while simultaneously improving the description of structural properties. Moreover, PBE provides a reliable account of the electronic structure. These findings clarify the role of correlation effects in CrOCl and identify PBE as a robust starting point for future ab initio studies of CrOCl-based materials.
\end{abstract}

\maketitle

 \section{Introduction}

Van der Waals (vdW) layered magnetic materials have recently garnered significant attention due to their unique magnetic properties and potential applications in spintronic devices as well as novel battery technologies \cite{Burch2018,Gibertini2019,Hao2022,Huang2017,Wang2020}. 
Among these materials, chromium oxychloride (CrOCl) has emerged as a promising vdW antiferromagnetic (AFM) semiconductor, not least because of its low exfoliation energy, which enables the easy fabrication of single layers \cite{Wang2020,Zhang2019b}.
%

\begin{figure*}
    \centering
    \includegraphics[width=\linewidth]{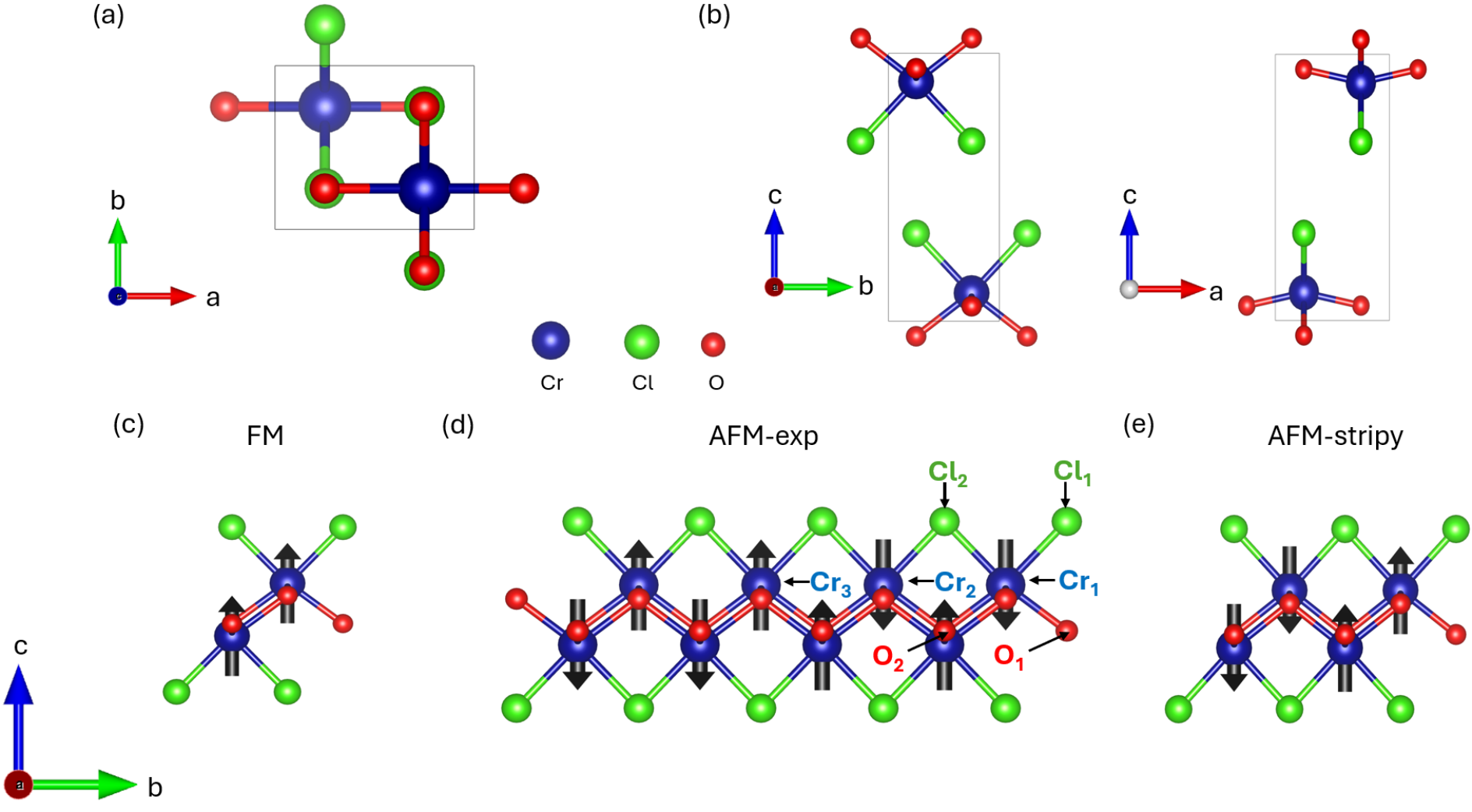}
    \caption{Structural illustration with (a) the top view ($ab$-plane) and (b) side view ($bc$- and $ac$-planes) of the orthorhombic CrOCl bulk system. Chromium (Cr) atoms are represented in blue, chlorine (Cl) atoms in green, and oxygen (O) atoms in red. Figures. (c)-(e) show various magnetic configurations represented by upward and downward arrows on the Cr magnetic atoms. 
    \label{fig:structure}
    }
\end{figure*}

Under ambient conditions, CrOCl adopts an orthorhombic structure with $Pmmn$ symmetry (space group 58), as depicted in Fig.\ \ref{fig:structure}.
This crystal structure is also shared by the TiOCl, VOCl and FeOCl systems.
Below the N\'eel temperature, $T_N=13.5$ K, CrOCl exhibits a complex intralayer four-fold AFM superstructure, accompanied by a unique monoclinic lattice distortion with $P2_1/m$ symmetry (space group 11)\cite{Schaller2023}, as shown in Fig.\ \ref{fig:structure} (d).

Ab-initio calculations of CrOCl often rely on the DFT+U formalism with the Perdew-Burke-Ernzerhof (PBE) functional \cite{Perdew1996} for exchange-correlation energy \cite{Jang2021,Gu2022,Zhang2023,Zhu2023a}.
However, several studies have shown that this approach tends to favor a ferromagnetic (FM) ground state for bulk CrOCl \cite{Miao2018,Zhang2019c,Nair2020,Qing2020,Xu2021,Zhu2023b}, in disagreement with the experimental AFM ground state. 
Jang \emph{et al.}\ and \cite{Jang2021} Zhu \emph{et al.}\cite{Zhu2023b} have independently pointed out that this may be an artifact of the commonly used Dudarev parametrization of the DFT+$U$ method \cite{Dudarev1998}. 
Jang \emph{et al.} explained this apparent failure of PBE+$U$ as the on-site exchange interaction being improperly described when a spin-polarized DFT functional is used, as suggested in Refs.\ \cite{Park2015,Chen2016}. Indeed, by using the unpolarized version of PBE, the experimental AFM state was recovered for the bulk\cite{Jang2021}. 
Another way to remedy this issue is to use the Liechtenstein formalism \cite{Liechtenstein1995} instead of the Dudarev parametrization \cite{Dudarev1998}, as it explicitly includes the on-site Hund's coupling parameter $J$, as suggested in Refs.~\cite{Gu2022,Zhu2023b}. Zhu \emph{et al.} \cite{Zhu2023b} showed that with a suitable choice of $J$, the correct AFM ground state is reproduced. 

In this work, we provide a new perspective on this matter by demonstrating that standard spin-polarized PBE (without $U$ or $J$ parameters) successfully reproduces the experimentally observed AFM ground state of bulk CrOCl.
At the same time, hybrid and meta-GGA functionals favor the FM state when applied at the room-temperature experimental lattice parameters.
Moreover, when combined with a simple van der Waals correction term, such as the D3 form \cite{Grimme2011}, PBE yields a better equilibrium volume and interatomic distances compared to DFT+$U$+D3 calculations. 
The PBE functional thus provides a very good description of structural, magnetic and electronic properties of the CrOCl system without the need for adjustable parameters.
These findings suggest that the inherent exchange splitting in PBE is entirely sufficient to capture the correct physics of CrOCl, challenging the conventional assumption that DFT+$U$ or hybrid functionals are required in this case. 

In Sec.~\ref{sec:magnetism} we report the FM-AFM energy balance with various functionals and basis sets. We then analyze the electronic structure in Sec.~\ref{sec:elstruct} and magnetic interactions in Sec.~\ref{sec:exchangeparameters} to clarify why the plain PBE functional performs unexpectedly well for CrOCl.

 \section{Computational Details}

Calculations with the Vienna Ab Initio Simulation Package (VASP) were based on the Projector Augmented Wave (PAW) method \cite{Kresse1999,Blochl1994}. 
In our calculations we treated the Cr ($3d$, $4s$, $3p$, $3s$), Cl ($2s$, $2p$), and O ($2s$, $2p$) states as valence.
A plane-wave cutoff energy of 650 eV was applied in all calculations, and the convergence criterion for total energy was set to 10$^{-6}$ eV. Optimized crystal structures were obtained with a force convergence threshold of 0.01 eV/\AA. 
For the k-point sampling, a Monkhorst-Pack k-point mesh of $19 \times 21 \times 13$ ($19 \times 5 \times 13$ for AFM supercell) was utilized for the DFT and DFT$+U$ calculations, while a $5 \times 7 \times 3$ ($5 \times 1 \times 3$ for AFM supercell) mesh was adopted for the hybrid functional calculations. We have also performed all-electron calculations with the Wien2k code \cite{wien2k}. using $R_{\text{MT}}=1.94, 2.09, 1.76$ a.u. for Cr, Cl and O, respectively. The basis set size was $R_{\mathrm{MT}}K_{\mathrm{max}} = 9.04$, $G_{\mathrm{max}} = 25\,\mathrm{Bohr}^{-1}$, and the $k$-mesh was $13 \times 4 \times 6$.
For the interatomic exchange-interaction (\(J_{ij}\)) calculations (hereafter denoted as \(J\)s), we first constructed a tight-binding Hamiltonian using maximally localized Wannier functions (MLWFs) generated via the \textit{Wannier90} package~\cite{Mostofi2008,Mostofi2014}. The MLWFs were derived from projector augmented-wave (PAW) calculations performed within the \textit{Quantum~ESPRESSO} code using the PBE-kjpaw\_psl.1.0.0.UPF pseudopotentials~\cite{Giannozzi2009}. The exchange interactions were subsequently evaluated using the Green’s-function method within magnetic force theory, as implemented in the \textit{TB2J} code~\cite{He2021}.
Visualization of crystal structures were created with the VESTA software \cite{vesta}.

\section{Results and Discussion}
\subsection{Magnetic ground state}\label{sec:magnetism}
\subsubsection{Fixed lattice calculations}\label{sec:fixedlattice}
\begin{figure}
    \centering
    \includegraphics[width=1\linewidth]{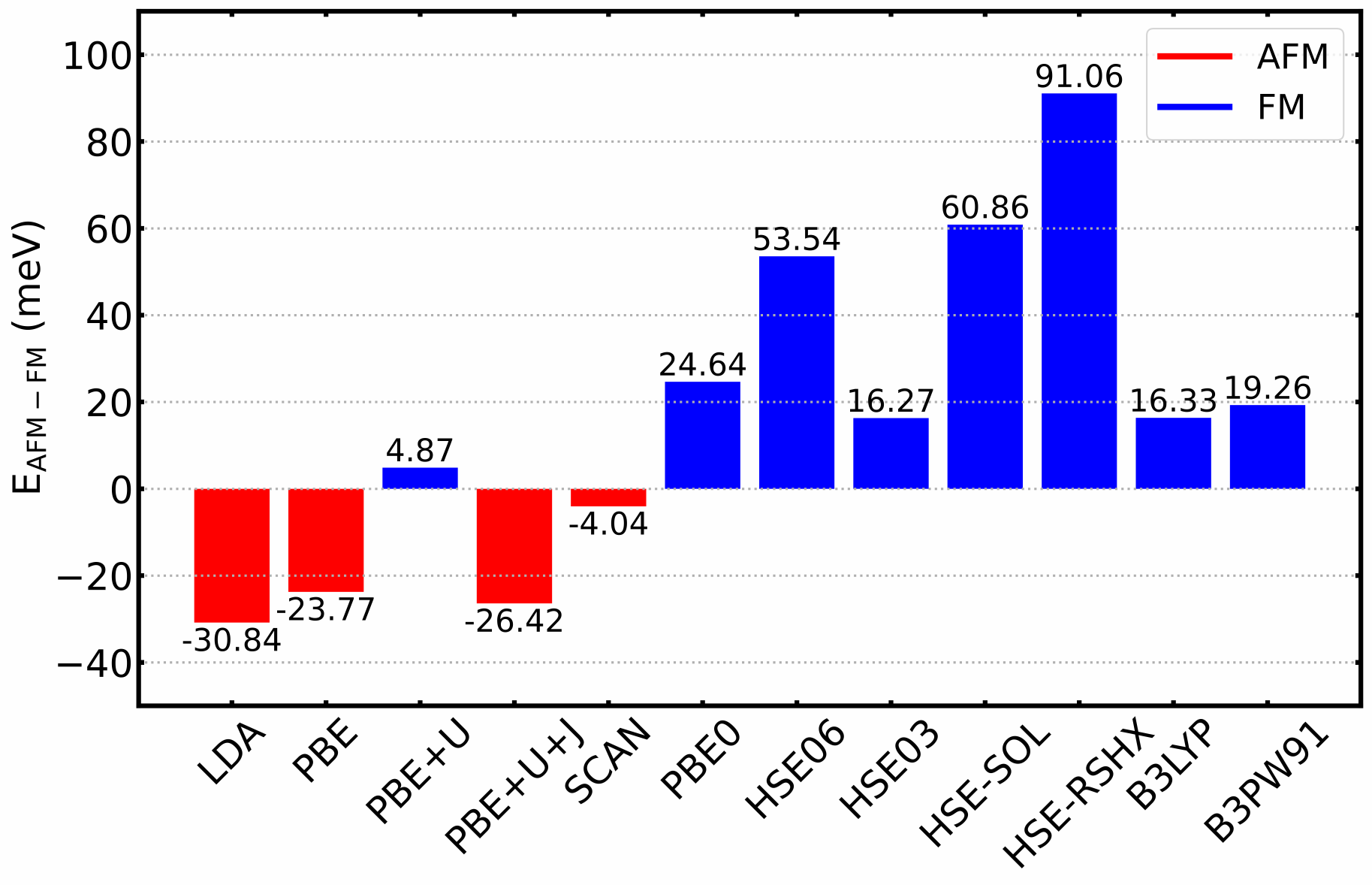}
    \caption{Energy differences (meV / unit cell) between the FM and experimental AFM configurations with various functionals for orthorhombic CrOCl with  experimental lattice coordinates. Positive values indicate the FM to be favored, while negative values indicate the AFM state is favored.
    \label{fig:barchart}
    }
\end{figure}

We first demonstrate how conflicting results are obtained for the magnetic ground state of CrOCl depending on the computational setup.
We adopt the the orthorhombic room-temperature experimental crystal structure with $a = 3.86 \, \text{\AA}$, $b = 3.18 \, \text{\AA}$, and $c = 7.69 \, \text{\AA}$ \cite{Norlund1975} and consider the energy difference between the FM and the experimental AFM configurations shown in Fig.\ \ref{fig:barchart}.
A wide range of exchange-correlation functionals were employed, from simple L(S)DA to DFT$+U$, the SCAN meta-GGA, and hybrid functionals.
We will refer to DFT+$U$ calculations with the Dudarev parametrization as PBE+U, and PBE+U+J for the Lichtenstein parametrization.
Following Ref.\ \cite{Zhu2023b}, we have set $U=3$ eV and $J=1.5$ eV, unless stated otherwise.

The results are summarized in Fig.\ \ref{fig:barchart}.
Indeed, the PBE$+U$+$J$ method correctly favors the experimental AFM state over FM order. 
In the Dudarev parametrization, the PBE$+U$ functional will erroneously favor a FM ground state, as pointed out in Refs.\ \cite{Jang2021,Zhu2023b}.

However, we also observe that with the PBE functional, we recover the correct experimental AFM state.
 Specifically, the experimental AFM state is 23.77 meV and 5.55 meV per unit cell lower in energy compared to the FM state and AFM-stripy configuration (Fig.\ \ref{fig:structure} (e)).
We also note that the even simpler LDA functional also favors the experimental AFM state by 29.21 meV.
These findings indicate that semi-local DFT functionals may accurately describe the magnetic interactions in CrOCl without the need to correct for the screened Coulomb interaction.

We also investigated the impact of spin-orbit coupling (SOC) on the magnetic ground state by performing non-collinear calculations with the \( c \) axis as the easy axis, as suggested in Ref.\ \cite{Angelkort2009}. 
The AFM state remained the stable configuration relative to the FM configuration even with the inclusion of SOC. 
Specifically, we found the AFM structure to be 22.71 meV lower in energy than the FM structure.
This is close to the calculated value of the $E_{\text{AFM-FM}} $ without SOC (23.77 meV), indicating that SOC is of minor importance.

Advancing to the meta-GGA level, we have chosen the strongly constrained and appropriately normed (SCAN) functional \cite{Sun2015}, which favors the experimental AFM state by 4.04 meV over the FM state.

We now turn to hybrid functionals, which have been used in high-throughput studies including CrOCl \cite{Mounet2018}.
We find that the screened HSE \cite{Heyd2003,Krukau2006} and RSHXPBE \cite{Gerber2005} functionals, as well as the unscreened B3LYP \cite{Stephens1994}, B3PW91 \cite{Becke1992}, PBE0 \cite{Perdew1996}, and SCAN0 \cite{Hui2016} functionals all favor the FM solution, as shown in Fig. 2. This negative result may indicate that optimizing the ion positions, or further extending the basis set, is required in order to reproduce the electronic structure. 
The application of hybrid functionals thus requires particular caution, as they may lead to an incorrect description of this system.

One study based on HSE06 suggested the stripy AFM configuration (Fig.\ \ref{fig:structure} (e)) to be lower in energy than FM, however the experimental AFM configuration was not considered \cite{Zhang2019}.
In our calculations, both the AFM-exp and AFM-stripy configurations are found to be higher in energy by 53.54 meV and 57.13 meV, respectively, compared to the FM state. Adopting the same lattice parameters as Ref. \cite{Wang2019}, we found the energy difference 40.87 meV, still favoring FM alignment over the AFM-stripy configuration, and leaving open questions about this discrepancy.
However, we do not pursue the performance of hybrid functionals further, as the main focus of this study is the performance of PBE in CrOCl.

To further identify reasons for apparent discrepancies between theory and experiment in CrOCl, we also tested all-electron full-potential LAPW calculations with the PBE functional. Using the experimental lattice constants in the LAPW calculations of Wien2k code \cite{Dewhurst2011} gives $E_{\text{AFM-FM}}  =  -27.9$  meV. These results further demonstrate the predictive accuracy of PBE in CrOCl which stabilize the correct AFM ground state.

\subsubsection{Structural Optimization}\label{sec:structure}

As demonstrated above, the PBE functional successfully reproduces the correct magnetic ground state at the fixed experimental lattice constants. 
We now perform a complete structural optimization.
To this end, we include the D3 method with the Becke-Johnson damping function \cite{Grimme2011} to account for vdW dispersion interactions.

\begin{table*}[ht]
\centering
\caption{Structural parameters obtained with PBE,  PBE$+U$ (with $+U$= 1 eV, 3 eV, 5 eV), and PBE$+U$+$J$ (with $+U$= 3 eV and $J$= 1.5 eV) including vdW interactions ( DFT-D3) compared with experiments at 3 K.
The labels Cr$_1$, Cr$_2$, Cr$_3$, Cr$_4$, and Cr$_5$ are explained in Fig.\ \ref{fig:structure} (d).
}
\begin{ruledtabular}
\begin{tabular}{ l c c c c c c c c c c c l}
 & $a$   & $b$   & $c$   & Volume & $c/a$ ratio &  $\alpha$ angle & Cr$_1$-Cr$_2$& Cr$_2$-Cr$_3$& Cr$_1$-O$_1$-Cr$_2$& Cr$_2$-O$_2$-Cr$_3$& Cr$_1$-Cl$_1$-Cr$_2$&Cr$_2$-Cl$_2$-Cr$_3$\\
 & \multicolumn{3}{c}{(\AA)} & (\AA$^3$)& & (deg)& (\AA)&(\AA)
&(deg)&(deg)& (deg)&(deg)\\  \hline
{Exp \cite{zhang2014temperature} } & 3.863& 3.178& 7.676& 94.230& 1.987& 90.070& 3.177& 3.161& 103.625& 102.847&   86.500&86.086\\  
 PBE & 3.872& 3.171& 7.738& 95.023& 1.998& 90.220& 
3.191& 3.150
& 103.637& 101.738&  86.853&85.840\\  
$U=1$ eV & 3.887& 3.190& 7.728& 95.834& 1.988& 90.106& 3.205& 
3.176
& 103.590& 102.244&  87.037&86.220\\  
$U=3$ eV & 3.913& 3.217& 7.770& 97.803& 1.986& 90.060& 
3.224& 3.210
& 103.449& 102.786&  86.918&86.515\\  
$U=5$ eV & 3.936& 3.242& 7.807& 99.614& 1.983& 90.039& 3.246& 
3.238
& 103.524& 103.144&  86.928&86.690\\
  PBE$+U$+$J$& 3.900& 3.192& 7.737& 96.338& 1.984& 90.080& 
3.202& 3.183
& 103.174& 102.355& 86.765&86.269\\  
\end{tabular}
\label{tab:structural_optimization}
 \end{ruledtabular}
\end{table*}

\begin{figure}
    \centering
      \subfigure[\label{fig:EV_LDA}]{\includegraphics[width=0.5\textwidth]{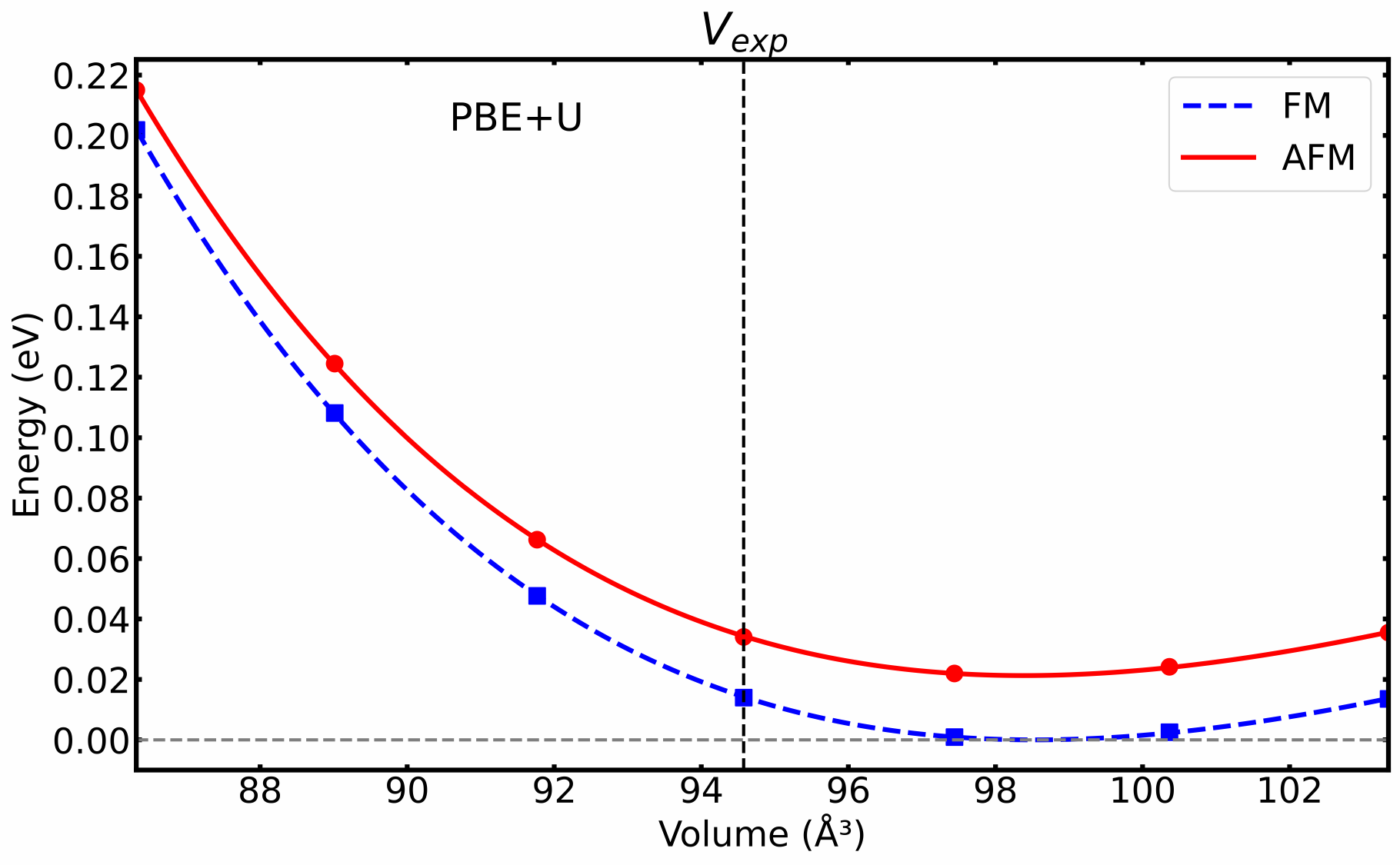} }
    \subfigure[\label{fig:EV_PBE}]{\includegraphics[width=0.5\textwidth]{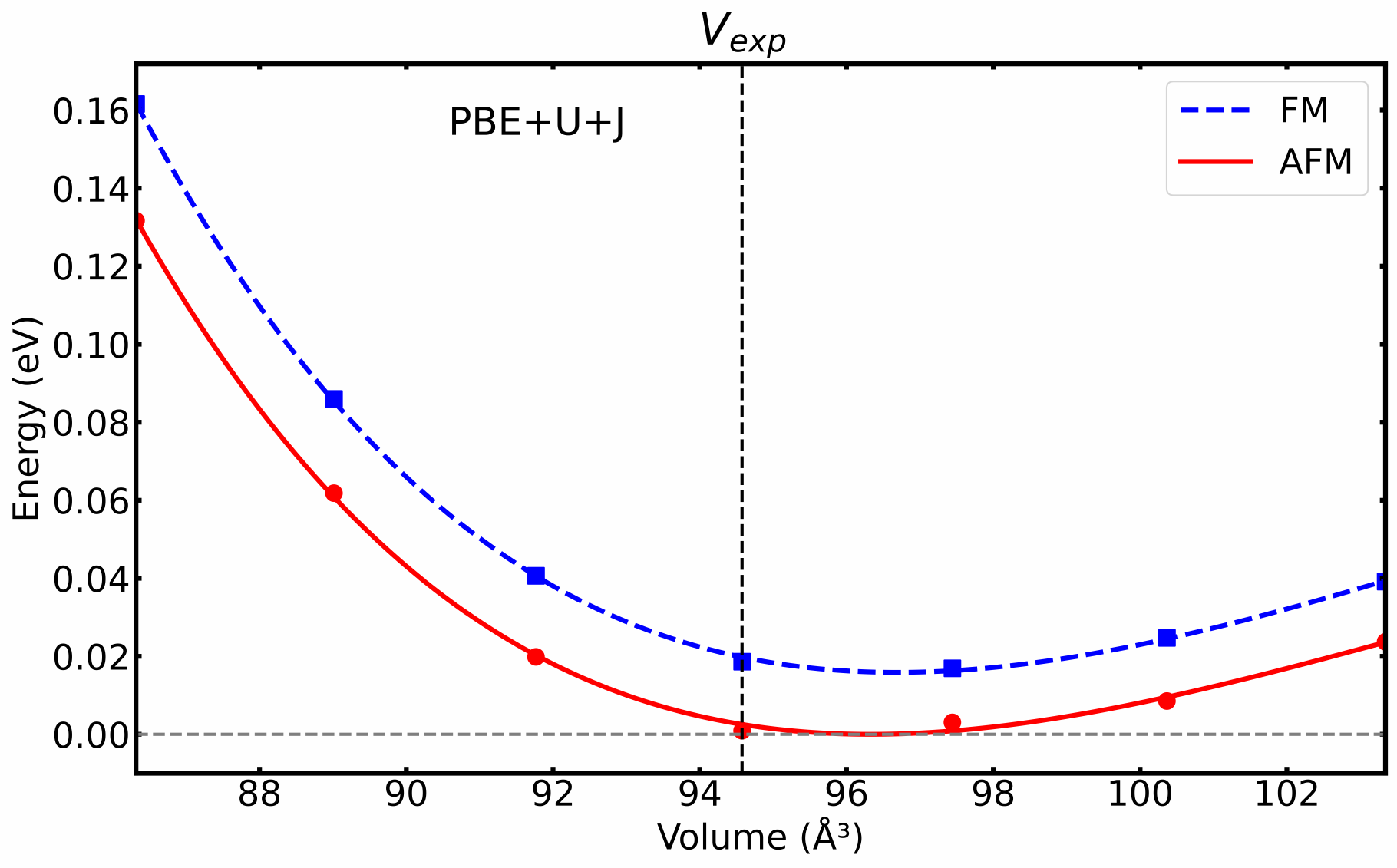} }
    \subfigure[\label{fig:Ev_PBEUJ}]{\includegraphics[width=0.5\textwidth]{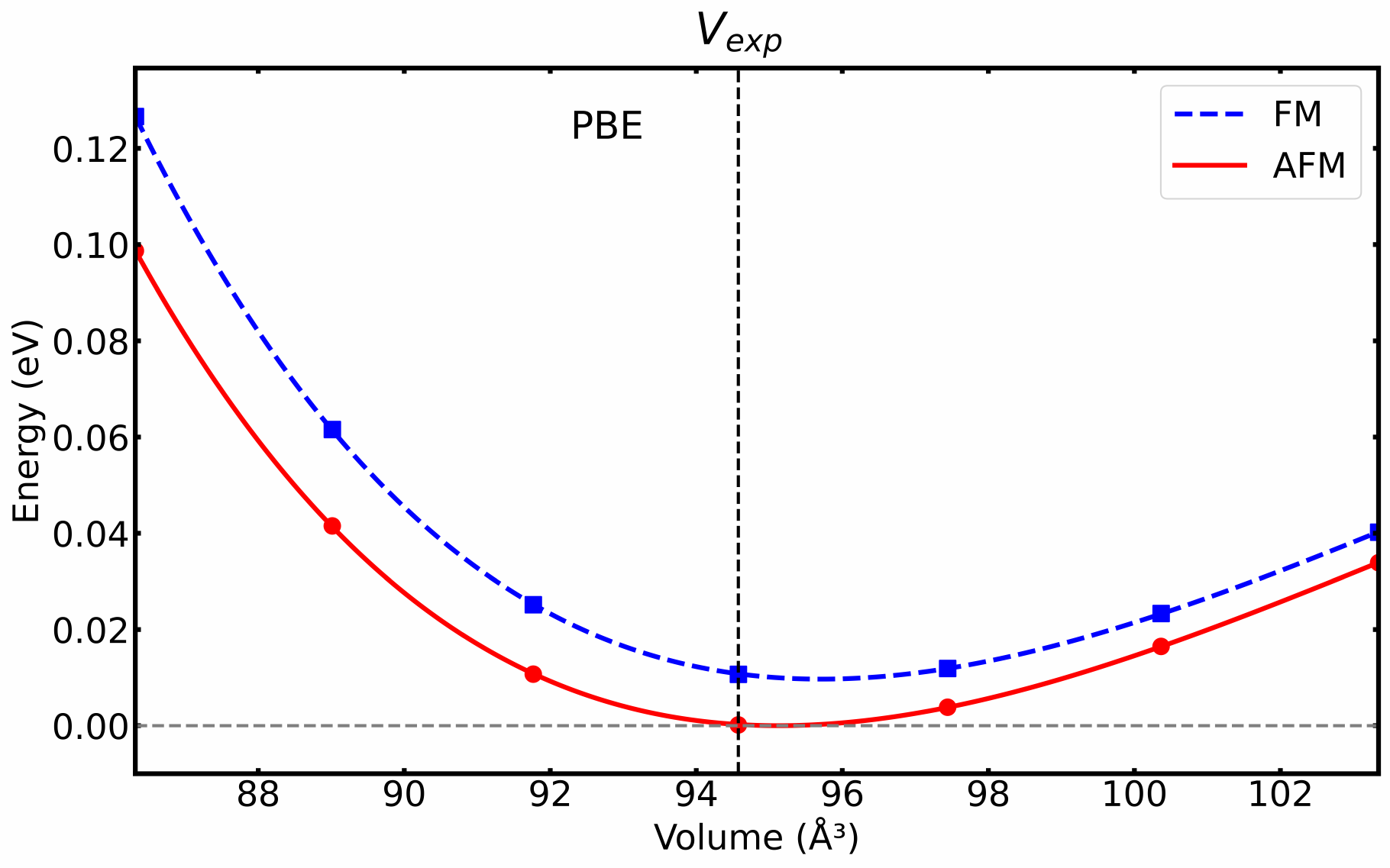} }
    \caption{Energy versus volume curves for antiferromagnetic (AFM) and ferromagnetic (FM) configurations obtained from (a) PBE$+U$, (b) PBE$+U$+$J$, and (c) PBE functionals. The equation of state (EoS) fits for the FM (dashed line) and AFM (solid line) configurations are fitted based on data points, with squares representing FM and circles representing AFM configurations. The energy is scaled to the minimum energy of the AFM structure. The horizontal dashed gray line indicates the zero energy reference, and the vertical dashed black line marks the experimental volume.}
    \label{fig:EV}
\end{figure}

Fig.\ \ref{fig:EV} shows total energy as a function of volume, as obtained with the PBE$+U$, PBE$+U$+$J$, and PBE functionals for the FM and the experimental AFM states. 
Indeed, PBE$+U$ with $U=3$ eV predicts the FM solution to be lower in energy than the AFM configuration in the entire volume range.
With the PBE$+U$+$J$ functional, the order is reversed, and the AFM solution is favored over FM. 
However, plain PBE correctly also recovers the AFM configuration as more favorable than FM for all considered volumes.
This shows that the failure of PBE+U and the success of PBE reported in Sec.\ \ref{sec:fixedlattice} are not due to the constrained geometry, but due to the inherent treatment of exchange-correlation effects.

We also note that PBE+U and PBE+U+J overestimates the equilibrium volume is overestimated relative to the experimental value. PBE$+U$+$J$ overestimates the volume by 2.2\%, as seen from Table \ref{tab:structural_optimization}, while the PBE equilibrium volume is only 0.8\% larger than the experimental volume measured at 3 K.

We also note that the L(S)DA functional favors the AFM configuration but significantly underestimates the equilibrium volume (see Supplementary Information \cite{suppmat}).

In Table \ref{tab:structural_optimization}, we also summarize the calculated equilibrium lattice parameters and selected interatomic distances in comparison with low-temperature experimental data.
The $a$ and $b$ lattice constants are only 0.2\% smaller than the experimental values with the PBE functional, while $c$ is 0.8\% larger.
The monoclinic distortion is exaggerated with PBE, reaching $90.22^\circ$.
Nevertheless, the nearest-neighbor Cr distances are in very good agreement with experimental values.

As $U$ is increased, the monoclinic $\alpha$ angle is reduced, improving agreement with experiment. 
However, the overestimation of the volume leads to overestimated nearest-neighbor Cr distances.

This trend is similar to the isostructural VOCl compound, where a larger \ueff\  tends to overestimate the volume but reduce the monoclinic distortion \cite{Ekholm2019}.

The inclusion of J in the PBE+$U$+$J$ calculation improves the accuracy of the structural parameters but still shows a larger deviation from experiments compared to the PBE results.

It is also worth noting that structural relaxation of the AFM-stripy configuration results in monoclinic distortion, although the angles deviate slightly, becoming sharper compared to the AFM-exp configuration.

Overall, the PBE functional performs very well, providing structural parameters that are generally closer to the experimental values compared to the PBE+$U$ and PBE+$U$+$J$ methods. The inclusion of U in the PBE+$U$ calculations leads to an increase in lattice parameters and unit cell volume, which deviate more from the experimental values.

\subsection{Electronic Structure}\label{sec:elstruct}
Having established the performance of the PBE functional in reproducing the experimental AFM state, we now analyze the electronic structure to see how it compares to other functionals. 

\subsubsection{Density of states}
\begin{figure*}
    \centering
    \includegraphics[width=\linewidth]{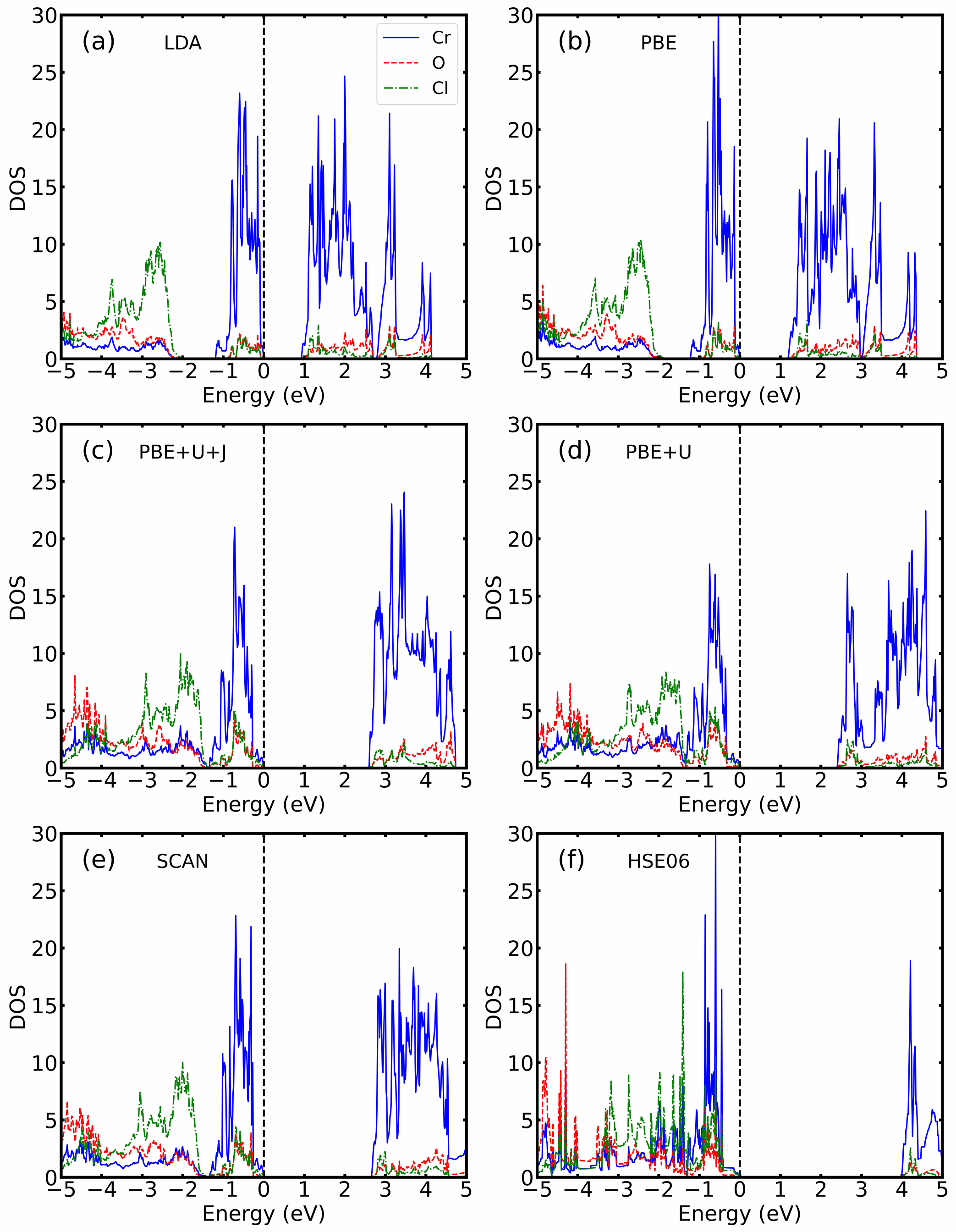}
    \caption{Site resolved density of states (DOS) for Cr, O, and Cl in various functionals: (a) LDA, (b) PBE, (c) PBE$+U$ ($U=3.0$ eV), (d) PBE$+U$+$J$ ($U=3.0$ eV and $J=1.5$ eV), (e) SCAN, and (f) HSE06. The solid blue line represents the DOS for Cr, the dashed red line for O, and the dot-dashed green line for Cl.}
    \label{fig:DOS}
\end{figure*}

In Fig.\ \ref{fig:DOS}, we present the calculated density of states (DOS) with various functionals for the same experimental AFM configuration and lattice constants.
Figs.\ \ref{fig:DOS} (a)--(b) shows the DOS calculated with the standard LDA and PBE functionals, which give very similar results for the valence states, with a band gap of approximately 0.88 eV for LDA and 1.05 eV for PBE. 
The experimental evidence for the size of the band gap is very limited. 
Co\"ic et al.\ \cite{Coic1981} performed optical absorption experiments, which indicated an onset energy just below 1.5 eV. The gap obtained with PBE is thus seen to be underestimated, which can be expected. Nevertheless the obtained gap is robust and clearly insulating.
However, it should also be noted that the measurements in Ref.\ \cite{Coic1981} were performed at 300 K, far above the N\'eel temperature.

We note that the the valence and conduction bands are dominated by Cr $d$ states. The former are in $d^3$ configuration, as the other $d$-states hybridize with the lower-lying Cl and O states $sp$-states, which are separated from the valence band by a gap. This is simlar to VOCl, which has $d^2$ configuration, but with a much narrower band width \cite{Amirabbasi2023}.

On the other hand, the PBE$+U$+$J$ functional, with $U=3.0$ and $J=1.5$, produces a larger band gap of about 2.45 eV. 
Notably, the separation between the valence $d$-states and the lower lying Cl and O states has been reduced to a pseudo gap at $-1.5$ eV.
This is in agreement with the PBE+U+J calculations reported in Ref.\ \cite{Jang2021}.
The PBE$+U$ functional, with \ueff$=3$ eV, produces a slightly smaller band gap of 2.2 eV.
However, the gap in the $-2$ eV to $-1.2$ eV region is absent, and the Cr valence states seem to hybridize with the Cl and O states.

With the SCAN functional, Fig.\ \ref{fig:DOS} (e), the band gap between valence and conduction states is close to that of PBE$+U$+$J$, while the separation between the Cr and Cl/O valence states is more pronounced. 
Among the hybrid functionals, we choose to show the HSE06 results in Fig.\ \ref{fig:DOS} (f).
The gap between the valence and conduction bands is almost 4 eV, and the Cr is heavily mixed with the ligand states.

It seems as if the ability of LDA/PBE to favor the AFM solution is connected to the opening of the gap between Cr valence band the rest of the occupied states, as the Cr--Cl/O--Cr superexchange mechanism may be influenced. 
Especially since the addition of a finite $J$ to the PBE+U functional is necessary to reproduce the AFM state, which also leads to a separation of the states.
The prononunced hybridization seen with HSE06 would then be connected with the large energy difference in favor of the FM state.

\begin{figure}
    \centering
    \subfigure[]{\includegraphics[width=0.5\linewidth]{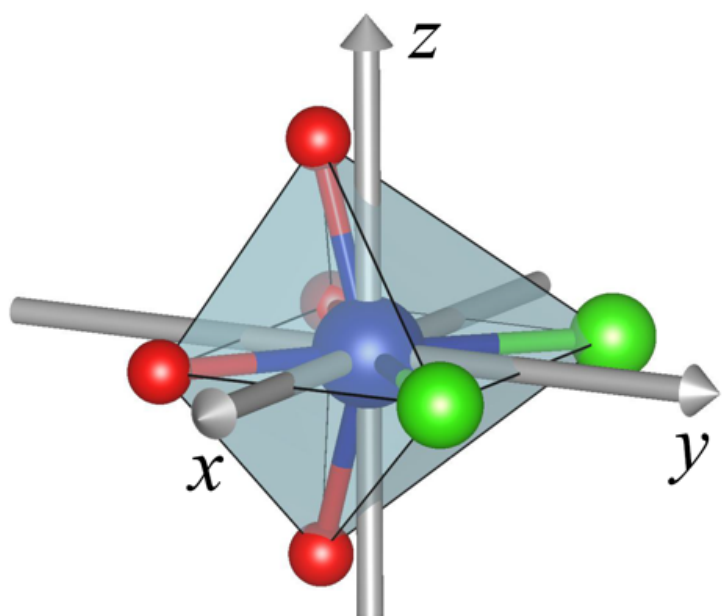}}
     \subfigure[]{\includegraphics[width=1\linewidth]{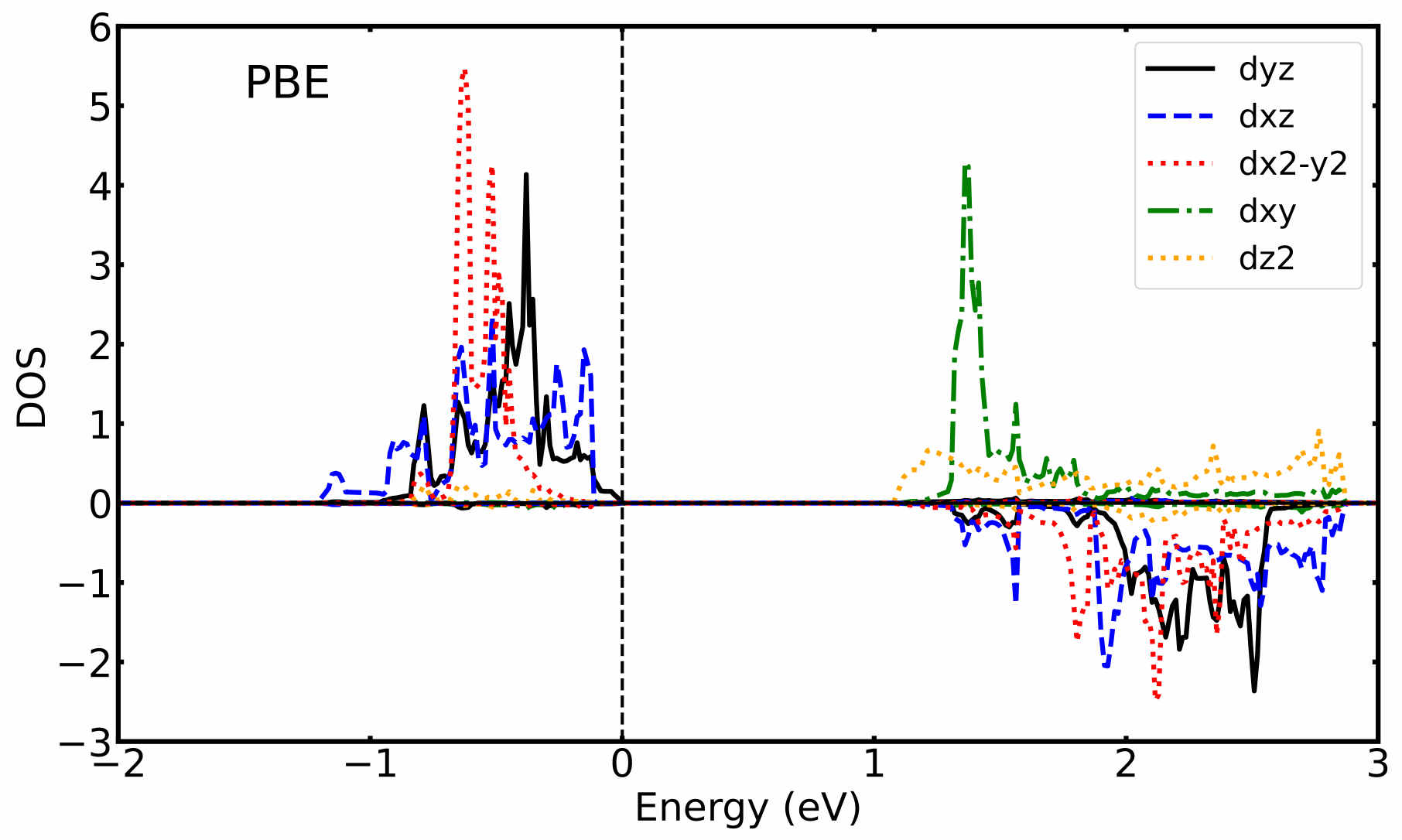}}
    \subfigure[]{\includegraphics[width=1\linewidth]{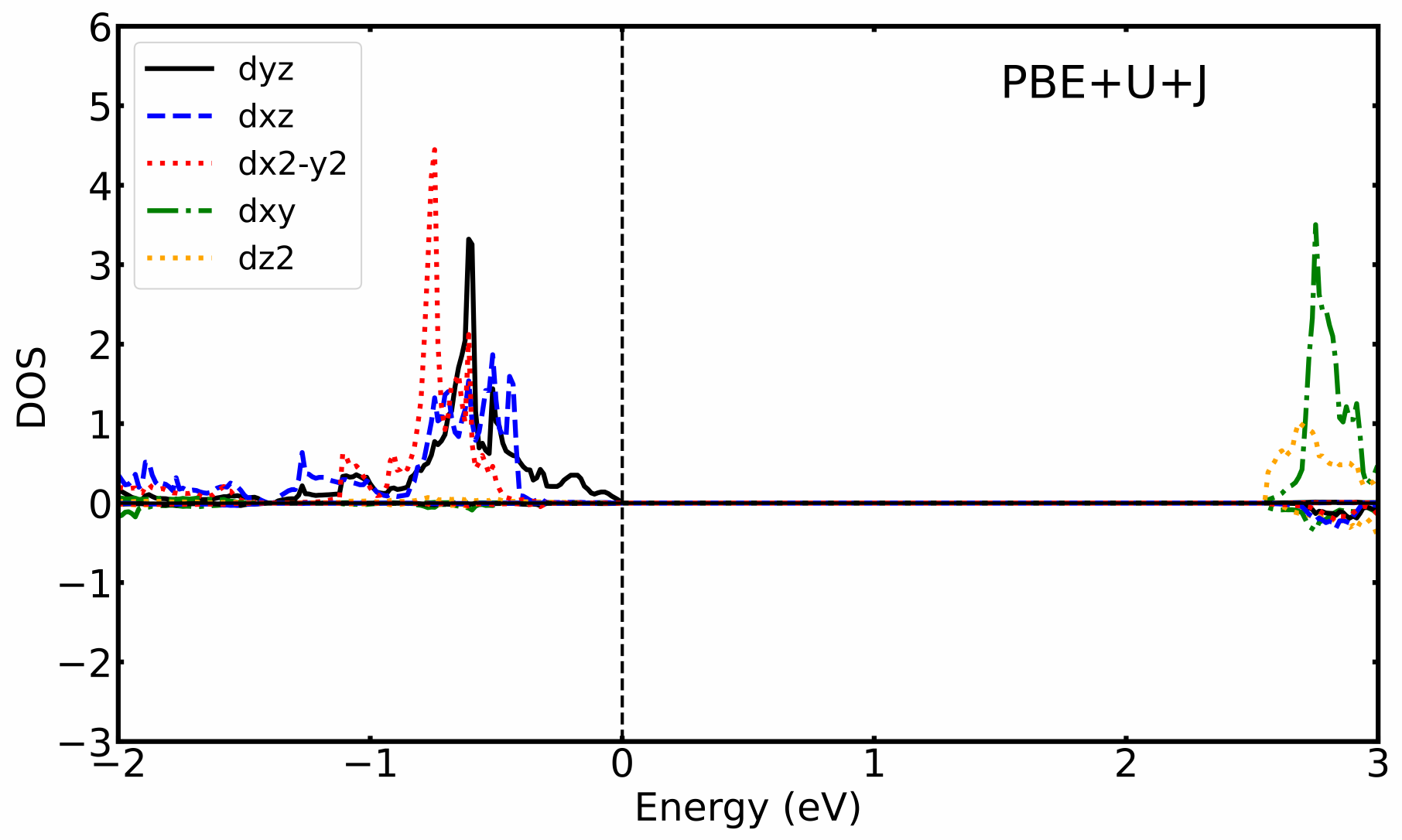}}
    \caption{(a) Distorted CrO$_4$Cl$_2$ octahedron with local $(x,y,z)$ coordinate frame, with $x$ along $\mathbf{b}$, $y$ along $\mathbf{c}$ and $z$ along $\mathbf{a}$. Spin resolved projected DOS obtained with (b) the PBE functional and (c) the PBE$+U$+$J$ functional.   \label{fig:3d-DOS}}
\end{figure}

Fig.\ \ref{fig:3d-DOS} shows the spin-resolved projected density of states (PDOS) of the valence $d$-orbitals with respect to a local $(x, y, z)$-system, where $x$ is along $\mathbf{b}$, $y$ is along $\mathbf{c}$, and $z$ is along $\mathbf{a}$.
Notably, the PDOS calculated with the PBE and PBE$+U$+$J$ functionals exhibits a similar orbital distribution.
The occupied $t_{2g}$ orbitals are comprised of the $d_{x^2-y^2}$, $d_{zx}$, and $d_{yz}$ spin-up orbitals.
The $t_{2g}^{\uparrow}$ states are thus filled in a $d^3$ high-spin configuration while the $e_g$ orbitals are empty, and the degeneracy is lifted by the crystal field of the strongly distorted CrO$_4$Cl$_2$ octahedron, producing a relatively simple crystal-field splitting and avoiding the
multiple complexity typical of strongly correlated oxides
\cite{ballhausen1962,zaanen1985}.

We note that compared to VOCl, the band width, $W$ of the valence states is much narrower in CrOCl. 
At the same time, they can in principle be described by the same $U$ and $J$ values, which means that the ratio $U/W$,  is smaller for CrOCl. 
This could be key to understanding the success of PBE: the system is only moderately correlated. It also interesting to note that unlike VOCl, the FM configuration will result in a very similar DOS, with a pronounced band gap (see \cite{suppmat}). This means that the mechanism for the opening of a gap is different in these isostructural systems.
In CrOCl, the mechanism is most likely due to the underlying band structure and not a Mott-type mechanism.

\subsubsection{Orbital splitting}

To quantify the orbital splitting seen in Fig.\ \ref{fig:3d-DOS}, we take the first-order moments:

\begin{equation}
    \varepsilon_n = \frac{\int E g_n(E) \, dE}{\int g_n(E) \, dE}
\end{equation}

where \( E \) represents the energy, and \( g_n(E) \) is the PDOS of orbital \( n \). In Table \ref{tab:orbital-splitting}, we present the orbital energies of majority spins relative to the lowest 3d orbital (\(d_{x^2-y^2}\)) energy for a Cr atom in the AFM configuration.

We observe that both PBE and PBE$+U+J$ yield consistent results for the occupied energy levels, but with a significant difference in unoccupied energy levels due to the larger band gap predicted by PBE$+U+J$. 
Within the occupied energy levels, the splitting energy between the \( d_{x^2-y^2} \) and \( d_{xz} \) orbitals is relatively small, while the splitting energy between the \( d_{x^2-y^2} \) and \( d_{yz} \) orbitals is larger. 

Comparing these results with previous calculations on VOCl reveals that the \( d_{yz} \) orbital becomes occupied in CrOCl, consistent with the addition of one more electron and the overall energy ordering. Additionally, the splitting between the \( d_{x^2-y^2} \) and \( d_{xz} \) orbitals is more pronounced in CrOCl than in VOCl, indicating that CrOCl displays stronger orbital differentiation within the occupied states.  It is also noteworthy that the energy levels of the \( d_{xy} \) and \( d_{z^2} \) orbitals in CrOCl are reversed in VOCl.

\begin{table}
 \caption{Orbital energies (\(\epsilon_n\)) of majority-spin \(d\)-orbitals relative to the lowest 3d orbital (\(d_{x^2-y^2}\)) for a single Cr atom in the AFM configuration using PBE and PBE$+U$+$J$ functionals, in comparison with previous calculations for VOCl in the AFM configuration \cite{Ekholm2019}.}
 \begin{ruledtabular}
     \begin{tabular}{l c c c c }
         & \(d_{xz}\) &  \(d_{yz}\)& \(d_{xy}\) & \(d_{z^2}\) \\ \hline
CrOCl (PBE)& 0.05 eV& 0.16 eV& 2.11 eV& 2.30 eV\\
CrOCl (PBE$+U$+$J$)& 0.04 eV& 0.14 eV& 3.64 eV& 3.72 eV\\
 VOCl & 0.19 eV& 2.44 eV& 3.39 eV&3.22 eV\\
\end{tabular}
\label{tab:orbital-splitting}
 \end{ruledtabular}
\end{table}

\subsection{Exchange Interactions}\label{sec:exchangeparameters}

\begin{figure}[hbt!]
    \centering
    \subfigure[\label{fig:Jij_AFM}]{\includegraphics[width=1\linewidth]{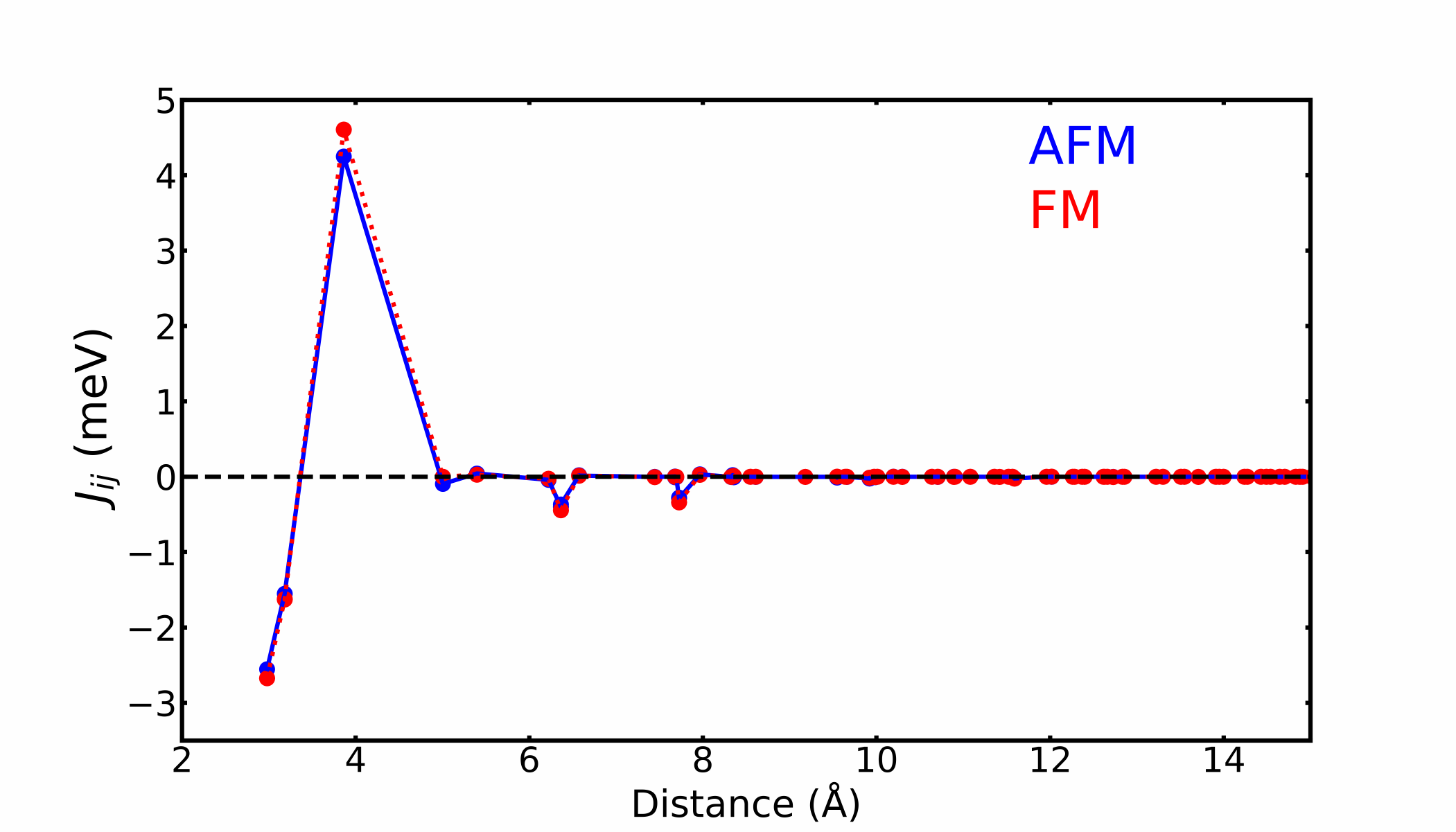}}
    \subfigure[\label{fig:Jij_illustration}]{\includegraphics[width=0.5\linewidth]{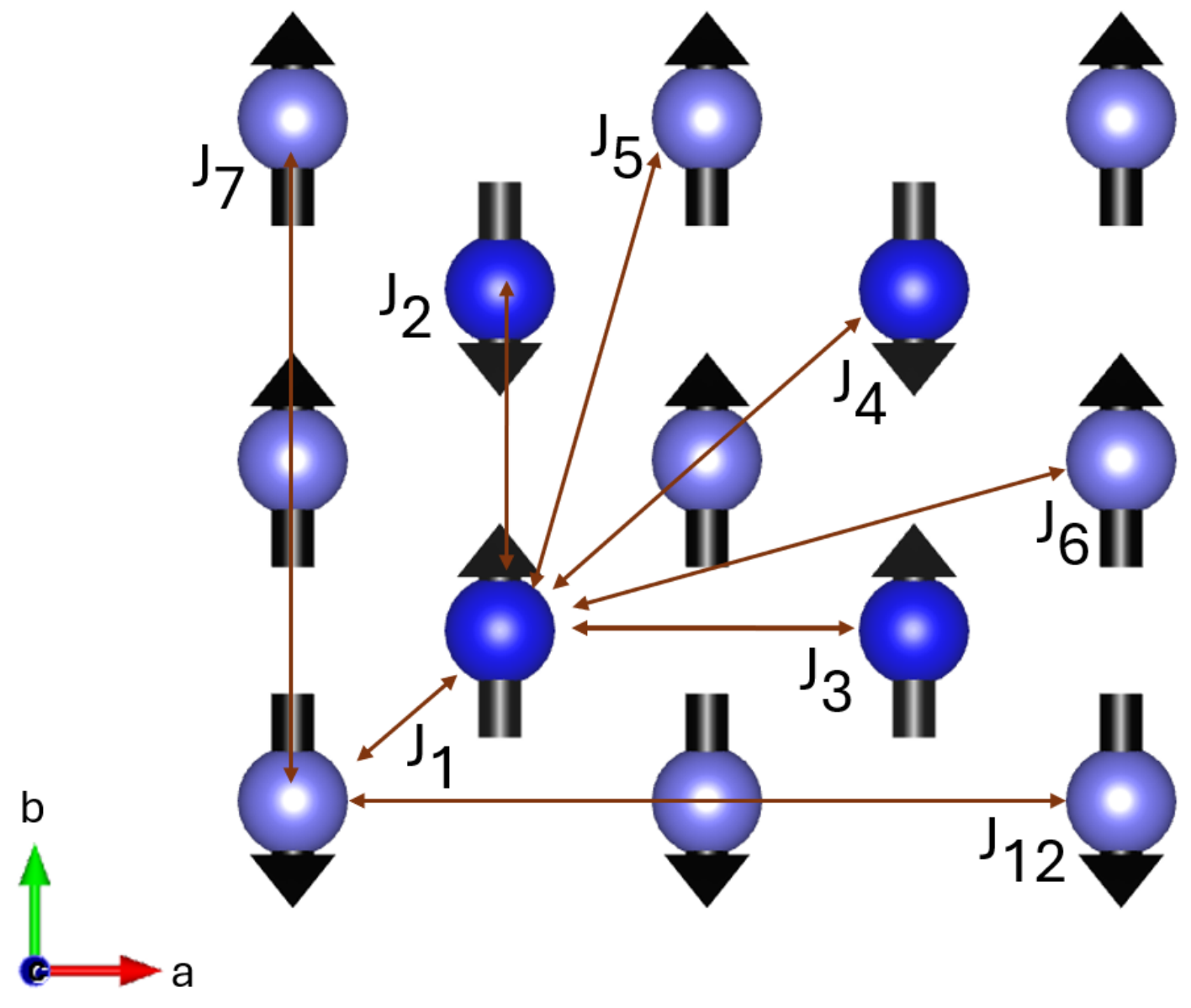}} 
    \caption{(a) Heisenberg exchange-parameters (J$_{ij}$) as a function of distance calculated from Fm and AFM reference states. (b) Illustration of magnetic exchange couplings for the experimental AFM configuration. The light and dark blue colors indicate Cr atoms with different $c$-coordinates, and arrows represent spin up and down.
    }   \label{fig:Jij}
\end{figure}

To compare the PBE picture of the AFM ground state with that of DFT+U \cite{Jang2021,Zhu2023b}, we have calculated the isotropic magnetic exchange interactions, $J_{ij}$, of the Heisenberg Hamiltonian:
\begin{equation}\label{eq:heisenberg_hamiltonian}
    H = -\sum_{i\neq j} J_{ij} \mathbf{\hat{e}}_i \cdot \mathbf{\hat{e}}_j \, ,
\end{equation}
using the magnetic force theorem \cite{liechtenstein1987local} and assuming orthorhombic crystal symmetry.
A positive sign of $J_{ij}$ indicates FM coupling, while a negative magnitude denotes AFM coupling. 
Exchange interaction matrices were calculated from both a FM reference state and the experimental AFM state. The results are very similar, as seen in Fig.\ \ref{fig:Jij_AFM}, which indicates that the exchange interactions are not sensitive to the local magnetic environment.

Numerical values of the most significant interactions are also presented in Table \ref{tab:exchange_constants}, and illustrated in \ref{fig:Jij_illustration}. $J_1$ and $J_2$ come out as AFM.
This is in line with the overall AFM alignment of spins along these vectors, although it should be noted that 1/4 of the Cr-Cr bonds along $\pm  \mathbf{a} \pm \mathbf{b} \pm z \mathbf{c}$ have FM alignment, and along $\pm \mathbf{b}$, each Cr atom has one FM and one AFM nearest neighbor, which introduces some frustration.

A key advantage of the Green’s function-based methodology over total energy mapping is its ability to disentangle the orbital contributions to the exchange parameters \cite{Yorulmaz2024}. 
For $J_1$ and $J_2$, the dominant contributions comes from direct $t_{2g}$-$t_{2g}$ hopping, which indeed favors AFM. The FM $t_{2g}$-$e_{g}$ hoppings corresponds to the Cr-O-Cr and Cr-Cl-Cr superexchange interactions.
$J_2$ is due to super-exchange through 
The $e_{g}$--$e_{g}$ interactions are an 1-2 orders of magnitude smaller.
With PBE+U, $J_1$ and $J_2$ turn out positive \cite{Zhu2023b}, which may be due to the increased hybridization with the ligand states. 
Just like PBE+U+J, the plain PBE functional gives an energy gap to the ligand states, which may be the reason why the direct exchange dominates superexchange.

There is also a strong FM $J_3$ interaction along $\mathbf{a}$, which enforces the strict FM alignment in this direction. 
$J_3$ is dominated by the FM $t_{2g}$-e$_{g}$ and AFM $t_{2g}$-t$_{2g}$ superexchange along the $158^\circ$ Cr-O-Cr path. It is interesting to note that PBE+U+J calculations yield a very small $J_3$-interaction, implying that these effects are cancelled out. 
The smaller insulating gap and larger hybridization gap in PBE may explain the preference of FM interaction, which naturally explains the stability of the experimental AFM configuration.  
All approaches also give a small, but important AFM $J_7$ interaction, which connects the antiparallel second-nearest neighbors at $\pm 2 \mathbf{b}$.

\begin{table}
 \caption{Isotropic exchange parameters J and their orbital decompositions from 1$^{\text{st}}$ NN to 7$^{\text{th}}$ NN, as well as the 12$^{\text{th}}$ NN interactions and their corresponding distances for AFM and FM configurations calculated with the PBE functional and PAW basis set. 
}
 \begin{ruledtabular}
     \begin{tabular}{l c c c c c c cl}
         AFM (meV)& J$_1$ & J$_2$ & J$_3$ & J$_4$ & J$_5$ & J$_6$ & J$_7$  & J$_{12}$\\ \hline
Total & -2.55& -1.55& 4.24& -0.10 & 0.05 & -0.05 & -0.39  & -0.29\\
 $t_{2g}$-$t_{2g}$ & -6.19& -4.73& -3.63& -0.02& 0.04& -0.18& -0.41&-0.51\\
 $t_{2g}$-$e_{g}$ & 3.87& 3.25& 7.70& -0.07& 0.01& 0.13& 0.06&0.24\\
 $e_{g}$-$e_{g}$ & -0.23& -0.06& -0.22& 0.01& 0.01& 0.01& -0.05&0.01\\

Distance (\AA) & 2.97 & 3.18 & 3.86 & 5.01 & 5.39 & 6.22 & 6.36  & 7.72\\
\end{tabular}
\label{tab:exchange_constants}
 \end{ruledtabular}
\end{table}

However, it is worth noting that the magnitude of the magnetic interactions is greater in the VOCl system than in the CrOCl system, which accounts for the lower Néel temperature in CrOCl (13.5 K) \cite{Amirabbasi2023,Das2023,Schaller2023} compared to that in the VOCl system (80 K) \cite{Wiedenmann1984}.

The PBE picture of the mechanism underlying the AFM--FM competition can now be understood in terms of the exchange parameters by writing the energy per Cr atom for FM order:
\begin{equation}
E_{\text{FM}} = -4 J_{1} - 2 J_{2}  - 2 J_{3} - 4 J_{4} - 4 J_{5} - 4 J_{6} - 2 J_{7} - 2J_{12} \, ,
\end{equation}
and for the AFM configuration:
\begin{equation}
E_{\text{AFM}} =  2J_1 -2J_3 - 2J_5 + J_6 + 2J_7 - 2J_{12} \, .
\end{equation}
The energy difference is then
\begin{eqnarray}\label{eq:E_AFM_FM}
E_{\text{AFM}}- E_{\text{FM}} 
    &\approx&  6J_1 +2J_2 + 4J_7 \, .
\end{eqnarray}
Using the parameters in Table \ref{tab:exchange_constants}, we obtain $E_{\text{AFM}}- E_{\text{FM}} $ within 1.8 meV of the DFT result. 
The $J_3$ interaction is seen canceled out from Eq.\ \eqref{eq:E_AFM_FM}, and the most crucial interactions are $J_1$, $J_2$ and $J_7$.
PBE thus favors the AFM state over FM because the $J_1$ and $J_2$ interactions are both AFM, just like in PBE+U+J.

\section{Summary and Conclusions}
We have shown that the plain PBE functional successfully recovers the experimental AFM ground state of CrOCl, while describing the structural properties on the same level, or better, than PBE+U+J calculations employing the Liechtenstein parametrization \cite{Liechtenstein1995}. 
Moreover, PBE is consistent with the current experimental picture, predicting antiferromagnetic order in CrOCl monolayers.
This is in contrast to standard PBE+U calculations with the Dudarev scheme \cite{Dudarev1998}, which incorrectly favor an FM ground state.

The success of spin-polarized PBE in describing bulk CrOCl is unexpected, as the addition of an on-site Coulomb correction typically is required in transition-metal oxides to stabilize an insulating state and obtain the correct magnetic properties.
Within PBE, there is a clear separation between the half-filled $t_{2g}$ states and the hybridized ligand $p$-metal $d$ states.
Adding the on-site Coulomb correction with PBE+U pushes the $t_{2g}$ states down into the $p-d$ manifold, which enhances FM superexchange between nearest-neighbor Cr atoms.
The PBE+U+J functional instead enhances the direct-exchange AFM interactions and thereby recovers the experimentally observed AFM ground state. 

A similar separation of states is obtained with the SCAN meta-GGA functional, which also favors the AFM configuration over ferromagnetism.
However, the addition of the Coulomb correction also leads to an overestimation of the lattice parameters.

In addition, the bandwidth of the occupied Cr $t_{2g}^{\uparrow}$ states is approximately 1.5~eV, which is roughly twice that found in VOCl, where an explicit Coulomb correction is required to open an insulating gap~\cite{Ekholm2019}. 
The larger bandwidth in CrOCl indicates that kinetic-energy effects are comparatively more important relative to on-site Coulomb interactions than in VOCl, which is consistent with the observation that PBE already yields an insulating solution. 
Notably, the insulating gap in CrOCl opens at the PBE level and persists even for ferromagnetic spin alignment, in contrast to VOCl where gap formation is more closely associated with antiferromagnetic order and enhanced electronic correlations. 
Taken together, these observations are consistent with an insulating state in CrOCl that is primarily governed by the underlying band structure rather than a Mott-type mechanism. 
This provides a plausible explanation for why PBE alone is sufficient to capture both the magnetic ground state and the structural properties of CrOCl.

Since CrOCl continues to attract interest for anisotropic optoelectronic, nonlinear optical, spin-dependent van der Waals heterostructures and high pressure studies\cite{Zhang2019b,Liu2024,Pawbake25,Schaller2023}, it is crucial to understand how to model its electronic and structural properties effectively.
The contradictory results obtained with different DFT+$U$ implementations have previously been discussed in the literature \cite{Jang2021,Zhu2023b}, where CrOCl was generally treated as a strongly correlated material. 
Our results provide additional insight into this issue by showing that exchange–correlation effects are already well described at the semi-local DFT level, and that the role of on-site Coulomb correlations in CrOCl may be more limited than commonly assumed.
As a continuation, it would be interesting to investigate the impact of the exchange-correlation functional on other physical properties, such as strain response and behavior under magnetic fields.

\begin{acknowledgments}
We acknowledge financial support from Olle Engkvists Stiftelse (Project No.~207-0582), the Swedish e-Science Research
Centre (SeRC), and Flair. The computations were enabled by resources provided by the National Academic Infrastructure for Supercomputing in Sweden (NAISS), partially funded by the Swedish Research Council through grant agreement no. 2022-06725, and by resources provided by the National Supercomputer Centre (NSC), funded by Link\"oping University.

\end{acknowledgments}

\bibliography{references}

\end{document}